\def\gsim{\lower.5ex\hbox{$\; \buildrel > \over \sim \;$}}
\def\lsim{\lower.5ex\hbox{$\; \buildrel < \over \sim \;$}}

\documentclass[cup5b]{caps}
\usepackage{graphicx}
\usepackage{amssymb}
\usepackage{ociwsymp3e}  

\HeadText{E. S. Perlman et al.}  

\def\plotone#1{\centering \leavevmode
\includegraphics[width=.95\columnwidth]{#1}}

\def\plotone#1{\centering \leavevmode
\includegraphics[width=.95\columnwidth]{#1}}

\begin{document}

\pagenumbering{arabic}

\author[]{E. S. PERLMAN$^{1}$, C. FRYE$^{1}$, H. EBELING$^{2}$, L. R.
JONES$^{3}$, C. A. SCHARF$^{4}$, D. HORNER$^{5}$
\\
(1) University of Maryland, Baltimore County,  Baltimore, MD, USA \\
(2) University of Hawaii, Honolulu, HI, USA \\
(3) University of Birmingham, Birmingham, UK \\
(4) Columbia University, New York, NY, USA \\
(5) Space Telescope Science Institute, Baltimore, MD, USA \\}

\chapter{The Evolution of Cluster Radio Galaxies at $0.5<z<1$}

\begin{abstract}

We are conducting a deep radio survey of a sample of 25 distant ($0.5<z<1$)
clusters of galaxies.    Here we present a progress report.    So far 17 of 25
clusters have been observed, to varying depths.  We have found 33 radio sources
within 0.3 Abell radii of the cluster center, of which 28 are likely associated
with cluster member galaxies. A comparison of the radio luminosity function of
these clusters with the results of two surveys at lower redshift, reveals no
evidence for strong evolution in the population of cluster radio galaxies at $z
\sim 0.65$.

\end{abstract}

\section{Introduction}

Rich clusters have long been known as the homes for powerful radio galaxies. 
Several of the first radio sources known lie in clusters (e.g., Perseus A,
Virgo A, Cygnus A), and surveys of nearby Abell clusters have discovered and
mapped many radio sources (e.g., Ledlow \& Owen 1996 and references therein). 
In low-redshift clusters, the radio sources have been found to be almost
exclusively of the lower power, edge-dimmed Fanaroff \& Riley (1974) class I
(FR I) variety, rather than more powerful, edge-brightened FR II sources.  It
was therefore a surprise when deep optical imaging of the fields surrounding $z
\sim 0.5-1$ radio sources found clusters around both FR Is and FR IIs (Prestage
\& Peacock 1988, Hill \& Lilly 1991, Ellingson et al. 1991; see also Harvanek
\& Stocke 2002).  Coupled with the then-prevalent notion that the X-ray
luminosity function of clusters evolved strongly at moderate redshifts (Henry
et al. 1992 and references therein), by the mid-1990s it appeared that 
at $z \sim 0.5$ radio sources in rich environments were responding to
the changes around them.

Since that time, our view of clusters and their radio sources has changed
significantly.  The most recent evidence on cluster environments seems to
indicate at most weak negative evolution since $z \sim 0.8$ (e.g., Jones et al.
1998, Borgani  et al. 1999, Rosati, Borgani \& Norman 2002 and references
therein), both in the X-ray luminosity function as well as in the X-ray
luminosity/temperature relation (Ettori, Tozzi \& Rosati 2003; Fairley et al.
2000).  This appears to contradict the earlier findings of the EMSS, albeit
mostly at 4-10 $\times$ lower X-ray luminosities given the greater depth but
smaller area of the more recent ROSAT-based surveys.  However, reanalyses of
the EMSS cluster data now also suggest weak or no evolution  in the cluster
X-ray luminosity function at higher luminosities out to $z=0.8$ (Nichol et al.
1997, Ellis \& Jones 2001, Lewis et al. 2002), although the error bars in the
high redshift bin are large.  It was therefore apparently  consistent with this
trend of finding little cosmological evolution in $z\sim 0.5$ cluster
environments when Stocke et al. (1999)  found no evolution in either the
luminosity function or morphology of radio sources in clusters, from a deep VLA
survey of 19 clusters at $0.3<z<0.83$ from the EMSS sample.

It was against this backdrop that we decided to use the WARPS sample (Scharf et
al. 1997, Perlman et al. 2002, Horner et al. in prep.) to re-examine the issue
of radio source evolution in clusters.   Due to its   depth (0.5--2.0 keV flux
limit of $\sim 6.5 \times 10^{-14} {\rm ~erg ~cm^{-2} ~s^{-1}}$) and sky
coverage ($\sim 70 {\rm ~deg^2}$) WARPS has a  much larger sample of high-$z$
clusters than existed only a few years ago.  At $z \gsim 0.5$, WARPS has 25
clusters, with a maximal redshift of 1.013.  By comparison, the EMSS in 1999
had only  6 clusters at $z>0.5$, with a maximal redshift of 0.829 (an
additional $z\sim 0.8$ cluster was added to the EMSS by Lewis et al. 2002; this
cluster is also part of WARPS and was in fact discovered by us.  See Ebeling et
al. 2000.).  With WARPS we therefore had the tool to not only attempt to
confirm the findings of Stocke et al., but also extend their result to
significantly greater lookback times.

In the ensuing sections, we assume a Hubble constant of $H_0=75 {\rm ~km
~s^{-1} ~Mpc^{-1}}$ and $q_0=0.1$ for consistency with Stocke et al. (1999)
and Ledlow \& Owen (1996).  

\section{Observations and Results}

To date, we have observed 17 of the 25 WARPS clusters at $z\gsim 0.5$. The
observed clusters range in redshift from $z=0.490$ to $z=0.92$, with a mean of
$\langle z \rangle = 0.65$.  But unfortunately the time awards for this project
did not allow us to observe the most distant cluster in the WARPS sample,
WARPJ1415.1+3612 at $z=1.013$ (Perlman et al. 2002).  The observations used the
VLA in either B or C configuration at a frequency of 1.4 GHz.  The observations
varied  in depth because the time allocations occurred at LST $\sim 19h-10h$;
however, they have averaged about 90 minutes, with multiple observations that
were well distributed in hour angle.  The typical $5 \sigma$ flux limit within
the central 2-3$'$  was 0.3-0.6 mJy.  

\begin{figure}
    \centering
    \plotone{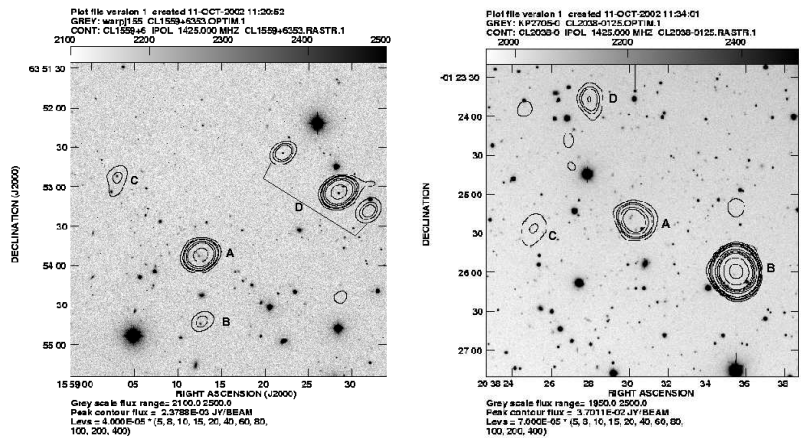}
    \caption[]{Overlays of VLA radio images (as contour maps) over optical
    images (greyscale) for two WARPS clusters of galaxies:  WARPJ1559.3+6353 
    (left, $z=0.81:$) and WARPJ2038.1$-$0125 (right, $z=0.679$).  Contours 
    are given at 3,5,8,12,16,24,32,48,64,128,256 $\times$ the rms noise.
    Both of these clusters have
    multiple radio sources located in their cores, as do most of the clusters 
    in our sample.  The radio sources are mostly associated with
    cluster member galaxies, but some are associated with galaxies in the 
    cluster's background or foreground (e.g., WARPJ1559.3+6353D, at left).}
\end{figure}

These observations detected 33 radio sources within 0.3 Abell radii ($A_c$) in
these 17 clusters.  This survey radius was chosen to match the work of Stocke
et al. (1999) and Ledlow \& Owen (1996).   Only two clusters (WARPJ0152$-$1357
and WARPJ0216$-$1747) did not have a radio source within $0.3 ~A_c$.  To decide
if these radio sources were associated with cluster member galaxies, we used
moderately deep $R$ or $I$ band images obtained at a variety of optical
telescopes,  and made radio-optical overlays.   We also compared to the results
of our optical spectroscopy (Perlman et al. 2002, Horner et al. in prep.). 
This allowed us to identify with reasonable certainty 21 of 33 radio sources
with galaxies that are likely cluster members.  The identifications of another
six radio sources with cluster member galaxies is less secure. Three radio
sources appear to be associated with optical blank fields, while  three radio
sources were identified with either foreground or background galaxies.   In
Figure 1 we show two sample overlays of optical and radio images.  

Once it is known which radio sources are associated with cluster members, it is
possible to compute their power at 1.4 GHz.  This was done assuming a spectral
index of 0.7 ($S_\nu \propto \nu^{-\alpha}$) for k-correction.  All but four of
the radio sources found within $0.3 ~A_c$ have radio power figures consistent
only with  those of FR I radio galaxies, which typically have $\log P({\rm 1.4
~GHz, ~W ~Hz^{-1}}) =  23-26$).  The remainder lie in the grey area occupied by
both FR I and FR II radio galaxies (the latter have typically $\log P({\rm 1.4
~GHz}) =  25-28$).  

The luminosity function for cluster radio galaxies is typically expressed in 
terms of the fraction of radio-loud, bright ellipticals in each bin of radio
power at 1.4 GHz.  Because we do not have complete spectroscopy for every 
galaxy within $0.3 A_c$, we estimated the number of bright galaxies within $0.3
A_c$ using the tight correlation between X-ray luminosity and galaxy counts
from Abramopoulos \& Ku (1983).  These figures were then multiplied
by a factor 2.2, in accordance with the standard form of the two-point
correlation function, to correct from a radius of 0.5 Mpc (used in Abramopoulos
\& Ku 1983) to $0.3 A_c$.  We then subtracted 20\% to account for the bright
cluster spirals and S0s, which do not contribute to the radio galaxy
population.   This method follows the analysis of Stocke et al. (1999) and
Ledlow \& Owen (1996).  The luminosity function achieved so far is shown in
Figure 2.

\begin{figure}
\plotone{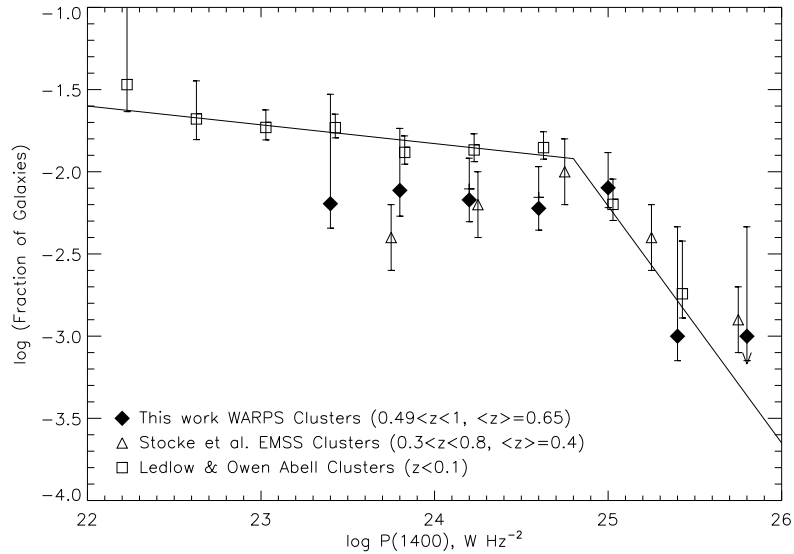}

\caption{The luminosity function of cluster
radio galaxies in the inner $0.3 A_c$.  The points shown in all but the highest
power bin represent our best estimates for radio source counterpart
identification, while the marks on the upper error bars represent the effect of
adding in blank field sources.  The point shown in the highest luminosity
bin represents a blank field source in one cluster (WARPJ2302.7$+$0844) which
is located very near the cluster center.  As shown, we do not see evidence for
strong  evolution in the population of cluster radio galaxies out to $\langle z
\rangle =0.65$.  The data may indicate weak negative evolution, but at
powers below the knee our results are likely affected by incompleteness (see
text). }

\end{figure}

\section{Discussion}

As can be seen in Figure 2, from our survey there is no evidence for strong
evolution of the luminosity function of cluster radio galaxies out to $\langle
z \rangle =0.65$, i.e., a lookback time 50\% greater than that which was
achievable with the EMSS (Stocke et al. 1999).  Thus our work has both
confirmed and extended to higher redshifts the result of Stocke et al. (1999),
which used a strategy similar to ours.  These results contrast with those
obtained through deep optical surveys of the environments around bright radio
sources (Prestage \& Peacock 1988, Hill \& Lilly 1991, Ellingson et al. 1991,
Harvanek \& Stocke 2002), which favor similar, moderately rich ($\sim$ Abell
class 0) environments for both FR I and FR II radio sources at z=0.5-1.    It
does, however, agree with the most recent results on the evolution of BL Lac
objects (Rector et al. 2000, Rector \& Stocke 2001), which now show that both
radio and X-ray selected samples have $\langle V/V_{max} \rangle \approx 0.5$. 

Our data, as well as those of Stocke et al. (1999) are suggestive of weak
negative evolution, as in both surveys the points at luminosities below the
knee lie below those seen at $z<0.1$.   However we believe this is due to
incompleteness.    The real problem with doing sensitive surveys of high-$z$
clusters lies in the dual issues of FR I sources' diffuse, edge dimmed
morphology (where the lobes tend to have $>90\% $ of the flux) and cosmological
$(1+z)^{-4}$ surface brightness dimming.  These two effects combine to
effectively lower the sensitivity of any survey to FR I - type sources. 
Unfortunately, since the effect is highly morphology dependent, it is difficult
to model.  Simulations were done for two prototype sources by Perlman \& Stocke
1993, but that work needs to be repeated on a much larger scale.  Whatever the
magnitude of the final effect of surface brightness dimming, we do know that it
would tend to move points towards artificially low densities in the FR I-only
bins.  Deeper VLA observations would help to address this problem. To firm up
either result (i.e, no or negative evolution), we also need better optical
observations, to firm up the identifications of radio sources with cluster
members.  As already discussed, for about 25\% of the radio sources found in
this survey, the nature of the optical counterpart (i.e., whether or not it is
associated with the cluster) is questionable.  In deriving the luminosity
function shown in Figure 2 we have used our best judgment; however, it is
possible that better optical imaging  could change the identifications assigned
to the more questionable cases and thus push the result towards one of negative
evolution.  Similar suggestions were made by Stocke et al. (1999) to explain
the fact that their lowest-luminosity point falls below the Ledlow \& Owen
(1996) curve.

The finding of weak negative or no evolution up to $z=0.65$ presents us with a
paradox, as not only does it conflict  obtained by optical surveys around radio
sources, it also appears inconsistent with that observed in most AGN (La Franca
\& Cristiani 1997, Ciliegi et al. 1995), which evolve positively, i.e., up to
$z\approx 2$ AGN were either more luminous or more numerous at higher redshifts
than in the present epoch. Stocke et al. (1999) suggested that this might be
explained by a higher density of FR Is in poor clusters.  This was prompted by
the evidence which then existed regarding the evolution of BL Lac objects,
which at that time  suggested negative evolution.  Our sample, which covers
X-ray luminosities 4 times lower than the EMSS at similar redshifts, does not
support this notion.  We cannot, however, comment on issues of morphology, as
we have no A-array data.

This apparent paradox is resolved by considering the likely association of FR
II radio galaxies with radio-loud quasars, which (like other quasars) appear to
evolve positively.  This is significant because the most recent evidence
suggests that FR IIs have lifetimes   $\lsim 10^{7-8}$ years, up to 5-10 times
shorter than those of FR Is, and that most radio sources do not evolve from  FR
II to FR I before they fade away (e.g., Parma et al. 2002).  If -- as seems
likely -- the evolution of the radio sources in clusters is intimately related
to the evolution of the cluster environment, one might expect the majority of
the evolution from FR II to FR I to take place at higher redshifts and yet
still find both types of radio sources in rich cluster environments. We can
then predict that deep radio surveys of  clusters at $z=1-2$ will find an
increasing prevalence of FR II-type objects.  This has not been seen in current
samples because very few (less than five) confirmed clusters of galaxies exist
at $z > 1$. Thus to tie in the evolution of the active fraction in clusters
with the evolution of the cluster environment, it is of key importance that
future samples of clusters which do reach $z\sim 2$ be surveyed with the VLA.

\end{document}